# Kagome electronic states in gradient-strained untwisted graphene bilayers

Zeyu Liu[1,2,3,#], Xianghua Kong[1,#,*], Zewen Wu[1,#], Linwei Zhou[1,2,3], Jingsi Qiao[4], Cong Wang[2,3], Shu Ping Lau[5] and Wei Ji[2,3,*]

[1]*College of Physics and Optoelectronic Engineering, Shenzhen University, Shenzhen 518060, China.*
[2]*Beijing Key Laboratory of Optoelectronic Functional Materials & Micro-nano Devices, Department of Physics, Renmin University of China, Beijing 100872, China*
[3]*Key Laboratory of Quantum State Construction and Manipulation (Ministry of Education), Renmin University of China, Beijing, 100872, China*
[4]*Advanced Research Institute of Multidisciplinary Sciences&School of Integrated Circuits and Electronics, Beijing Institute of Technology, Beijing 100081, China*
[5]*Department of Applied Physics, Hong Kong Polytechnic University, Hung Hom, Kowloon, Hong Kong, P. R. China.*
*Emails: kongxianghuaphysics@szu.edu.cn (X.K.), wji@ruc.edu.cn (W.J.),

**ABSTRACT:** Moiré superlattices in twisted homo-bilayers have revealed exotic electronic states, including unconventional superconductivity and correlated insulating phases. However, their fabrication process often introduces moiré disorders, hindering reproducibility and experimental control. Here, we propose an alternative approach using gradient strain to construct moiré superlattices in untwisted bilayer graphene (gs-BLG). Through force-field and first-principles calculations, we show that gs-BLG exhibits kagome-like interlayer-spacing distributions and strain-tunable kagome electronic bands. The competition between interlayer coupling and in-plane strain relaxation leads to distinct structural deformations, giving rise to three forms of diatomic kagome lattices: subtle, pronounced, and distorted. Kagome electronic bands are identified near the Fermi level in their band structures. Modulating strain gradients enables tailoring bandwidths and signs of hopping parameters of these kagome bands, providing a versatile platform for studying exotic electronic phases. Our findings establish gradient strain as an alternative to twist engineering, opening an avenue for exploring emergent electronic phases in graphene-based systems.

**Introduction**

Two-dimensional moiré superlattices can be formed by stacking two-dimensional (2D) materials, either homogeneous or heterogeneous, using strategies such as twisting or mismatched lattices. Numerous combinations of lattice-mismatched 2D hetero-bilayers and tunable twist angles diversify the periodicity and symmetry of moiré superlattices[1]. Electronic flat bands and their resulting strong correlation effects were observed in twisted bilayer graphene[2-4] and transition metal chalcogenides[5,6]. A wide spectrum of resulting intriguing states, including unconventional superconductivity[7-9], correlated insulating states[10-13], Wigner crystals[14] and rich topological features[15-18] were found in twisted bilayers. Moreover, the slowly varying moiré potential in moiré superlattices significantly promotes the formation of moiré excitons, including both inter- and intra-layer excitons, as well as moiré trions[19-21].

However, even slight structural perturbations can significantly vary the electronic properties of these moiré superlattice[22-24]. For instance, superconducting states observed in magic-angle twisted bilayer graphene may vanish even under fluctuations of 0.1° for the twisting angle or 1 nm for the superlattice periodicity[25,26]. Additionally, the introduction of twisting in most experimental assembly techniques inevitably induces strain fluctuations, referred to as moiré disorders[22,27-29]. The twisting-induced moiré disorders severely limit the reproducibility of experiments and thus the alignment with theoretical findings[30-34]. Moiré disorders are much less pronounced in untwisted moiré superlattices where exotic electronic states are still present. For instance, fractional Chern insulating states were predicted in untwisted but periodically strained monolayer graphene[35]. Moreover, a one-dimensional flat band was experimentally observed in an untwisted Bi(110)/SnSe(001) heterostructure[36]. Graphene is the strongest 2D material ever measured, which can endure a reversible tensile strain up to 25%[37-40], showing its potential for forming moiré superlattices implemented by strain. By applying a bubble structure, bilayer graphene exhibits a strain difference of approximately 2 % between the two layers[41-45].

Inspired by experimental observations, here, we theoretically construct gradient-strained bilayer graphene (gs-BLG) moiré superlattices by introducing an in-plane strain difference in untwisted bilayer graphene. Using force-field structural relaxations and first-principles density functional theory (DFT) calculations, we reveal kagome-like interlayer-spacing patterns and resulting kagome electronic bands (flat bands and flattened Dirac bands) near the Fermi level. We first present the structural models of gs-BLGs, highlighting differences among four local stacking configurations. Structural relaxations reveal distinct interlayer-spacing maps forming subtle, pronounced, and distorted diatomic kagome lattices. Our DFT calculations verify the emergence of kagome bands, with wavefunction visualizations validating their real-space distribution. Furthermore, we demonstrate that strain gradients can be used to tune kagome band dispersion, offering an approach to engineering electronic properties. Finally, we establish a direct correlation between out-of-plane corrugation and kagome patterns, demonstrating gradient strain as an effective strategy for constructing moiré superlattices and realizing novel electronic states.

## Results

### Geometric Structure of Gradient-Strained Bilayer Graphene

A gs-BLG moiré superlattice could be constructed by stacking a $3N \times 3N$ and a $(3N-1) \times (3N-1)$ graphene monolayer supercells together. A series of gs-BLG moiré superlattices can be constructed in varying strain rates by adjusting the supercell size ($N$) (Fig. S1). Figure 1a shows a gs-BLG model, without structural relaxations, for $N=12$, in which the gs-BLG consists of a $35 \times 35$ (gray, top-layer) and a $36 \times 36$ (red, bottom-layer) graphene supercell. Its electronic bandstructures and real-space distributions of wavefunctions are briefly discussed in Fig. S2. Each gs-BLG moiré superlattice contains four distinct high-symmetry stacking configurations, namely AA, AB, BA and SP, as highlighted in Fig. 1a and detail discussed in Fig. 1b. In the AA stacking (blue in Fig. 1), each atom of the top-layer is positioned directly over a carbon atom of the bottom-layer, a half-unit-cell lateral shift results in the AB (violet) or BA (orange) stacking. Moreover, there is a transition stacking configuration between AB and BA, referred to as the SP (saddle point) stacking (green). The optimal layer spacings and relative energies of these stacking configurations, together with their atomic models, are depicted in Fig. 1b.

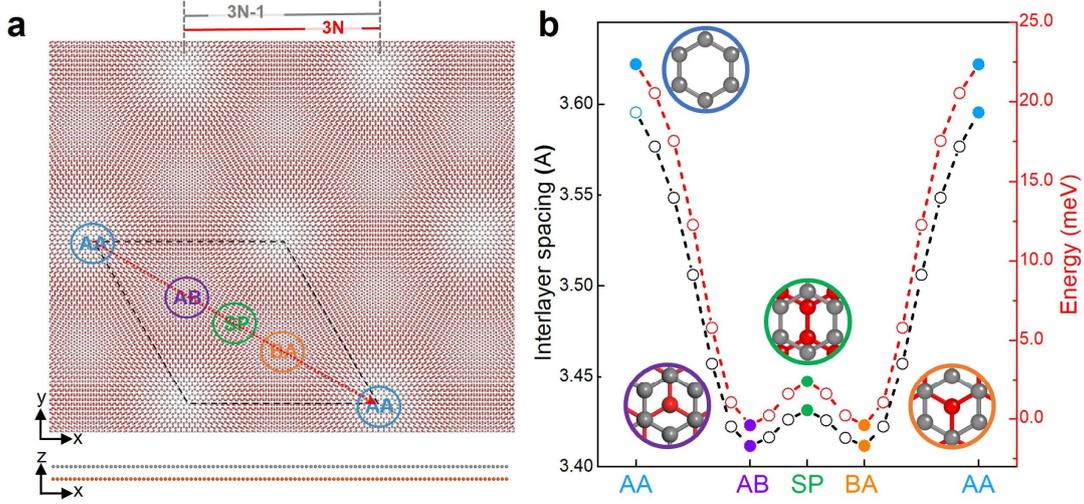

Figure 1 (a) Top- (upper panel) and side-views (lower panel) of atomic structures of an in-plane gradient strained bilayer graphene moiré superlattice. The top-layer (in grey) consists of $3N$-1 × $3N$-1 (here, $N$=12) graphene unit-cells and the bottom-layer has $3N$ × $3N$ graphene unit-cells. The top- and bottom-layer are subjected to potential strains to ensure they have the same in-plane lattice constants for the bilayer supercell. The supercell is marked by the black dashed diamond. Within the diamond, four high-symmetry stacking orders are denoted by differently colored circles, with inset texts showing their labels. (b) Optimal interlayer spacing and relative energy profile of four high-symmetry stacking orders along the red dotted arrow shown in (a). Atomic models are presented for these four stacking orders inside the circles with associated colors.

In gs-BLG models, the top and bottom layers share the same superlattice constant. The in-plane strains of the two layers can take three primary combinations, determined by the supercell lattice constant: 1) the top-layer stretches to conform the non-strained bottom-layer (SN-gs-BLG); 2) the top-(bottom) layer stretches (compresses) to accommodate an averaged lattice constant (SC-gs-BLG); 3) the bottom-layer compresses to match the non-strained top-layer (NC-gs-BLG). Two additional combinations, where both layers are simultaneously stretched (SS-gs-BLG) or compressed (CC-gs-BLG), also exist and show geometry and electronic structures similar to SN- and NC-gs-BLGs, respectively. Therefore, this work focuses on the first three combinations.

Figures 2(a-c) show the fully relaxed atomic structures of these three gs-BLGs for $N$=12, exhibiting distinct in-plane strain values: 2.86 % stretching for the top-layer in SN-gs-BLG (case 1, Fig. 2a), ±1.44 % averaged strain for SC-gs-BLG (case 2, Fig. 2b),

and -2.78 % compression for the bottom-layer in NC-gs-BLG (case 3, Fig. 2c). Relaxed SN- and SC-gs-BLGs structures feature patterns combining hexagonal (AB and BA regions, violet and orange circles) and triangular (AA regions, blue circles) lattices, as depicted in Figs. 2a and 2b. However, NC-gs-BLG displays a three-pointed star pattern centered at the AA stacking regions (blue circles). The SP stacking regions (green circles) shifted toward the AA regions, causing the original AB and BA regions to rotate and expand, maximizing their areas (Fig. 2c).

These structural variations are lined to out-of-plane distortions, represented with out-of-plane local atomic displacement ($\Delta L_{\text{Oop}}$), which establish distinct interlayer-spacing ($\Delta d$) patterns. Figures 2(d-f) present the maps of interlayer spacing for the three gs-BLGs, highlighting the relative out-of-plane distortions. While interlayer spacing is closely tied to stacking order, different in-plane strain combinations significantly reshape spatial distribution patterns. Regions with the smallest interlayer spacing are represented in dark red in Figs. 2(d–f), form subtle diatomic (SN-gs-BLG, Fig. 2d), pronounced diatomic (SC-gs-BLG, Fig. 2e), and distorted diatomic (NC-gs-BLG, Fig. 2f) kagome lattices, delineated by green dashed triangles and lines. Additional visualizations and discussion of these structures and out-of-plane structural corrugations are provided in Fig. S3.

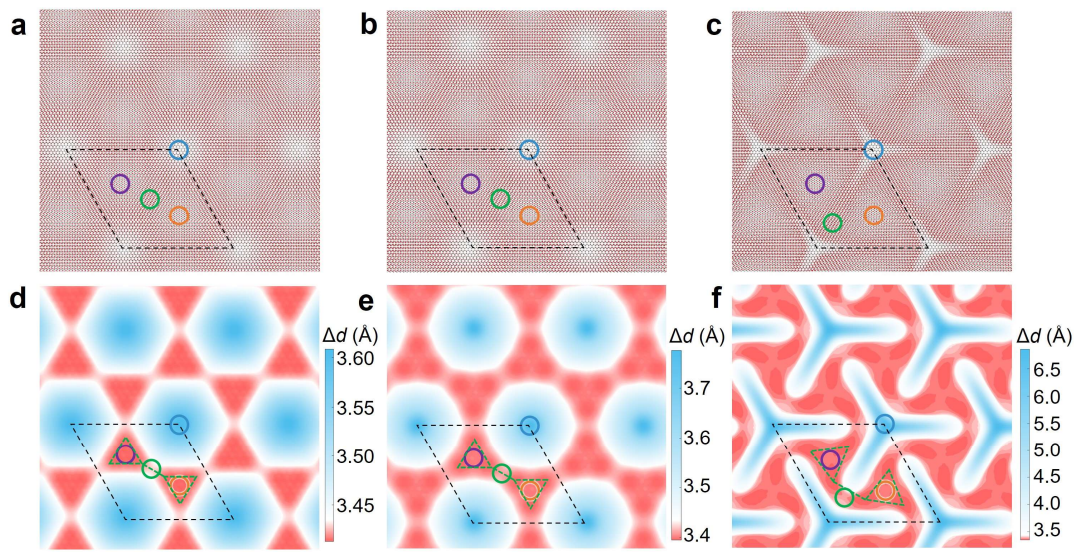

Figure 2. Atomic structure of gradient-strain bilayer graphene. (a) Top views of the atomic structures for: SN-gs-BLG under 2.86% stretched strain applied to the top-layer, (b) SC-gs-BLG under ±1.44%

stretched (compressive) strain applied to the top (bottom)-layer, and (c) NC-gs-BLG under -2.78% compressive strain applied to the bottom-layer. (d-f) Projections of the corresponding interlayer spacing of the three, and the white in color bar corresponds to the optimal interlayer spacing of the SP stacking. Black dashed diamonds and differently colored circles represent the supercell and the four stacking orders in the same scheme used in Fig. 1a.

Interlayer-spacing patterns arise from a competition between in-plane elastic (strain) energy and interlayer coupling energy. Both SN- and SC-gs-BLGs exhibit similar interlayer-spacing patterns (Fig. 2d and 2e), primarily ascribed to biaxial in-plane stretching. While out-of-plane deformation is minimal under in-plane stretched strains, the distribution of in-plane deformation depends on interlayer interactions. In strongly coupled AB and BA stacking domains (red in Fig. 2d and 2e), local lattice mismatch is minimized to preserve interlayer coupling strength. Conversely, weakly coupled AA stacking domains (in blue) concentrate most in-plane deformation, efficiently relieving stress with minimal interlayer interaction breakdown cost. This redistribution of structural deformation also leads to small out-of-plane corrugations (< 0.4 Å) around the AA regions facilitating strain release. Figure S4 illustrates the in-plane distortion [local atomic displacements ($\Delta L_{IP}$)] and out-of-plane distortion distributions, supporting this analysis.

Under pure compression in NC-gs-BLG, the in-plane elastic (strain) energy increases substantially, causing the interlayer coupling energy of the original AB and BA domains to be insufficient to balance the heightened strain energy. Thus, out-of-plane distortions become significantly amplified (see Fig. S4), more effectively relieving compressive stress. The interlayer spacing in the AA regions increases from 3.4 Å to over 6.5 Å (Fig. 2f). Moreover, strongly coupled AB and BA domains (red in Fig. 2f) expand their areas through in-plane rotations, namely AB domains rotate counterclockwise, while BA domains rotate clockwise, further lowering the total energy of compressed NC-gs-BLG. These rotations drive the SP stacking regions (green circles) to converge, from the midpoints of AB and BA domains, toward the AA regions (Figs. 2c and 2f). A distorted diatomic kagome lattice forms by connecting regions of small interlayer spacing (dark red regions), as delineated by dark green dashed triangles and lines in Fig. 2f.

**Electronic Properties of Gradient-Strained Bilayer Graphene**

Given the formation of three diatomic kagome lattices, kagome bands are anticipated in the band structures of these gs-BLGs, as confirmed by our DFT calculations. Figure 3a presents the layer-projected band structure of SN-gs-BLG, highlighting kagome bands (green dashed lines) characterized by a Dirac point (DP1) and a flat band (FB1). A lower-energy nominal FB (residing near the Fermi level at the Gamma point) appears obscured due to perturbations from neighboring bands, which, together with bands FB1 and DP1, exhibit partial characteristics of diatomic kagome bands. The diatomic kagome band features become pronounced in SC-gs-BLG, aligning with the interlayer-spacing map, as illustrated in Fig. 3b. The previously observed DP1 and FB1(in green) persist, while new DP2 and FB2 bands (in brown) emerge, exhibiting quintessential electronic band features of a diatomic kagome lattice[46]. Figure 3c presents the band structures of NC-gs-BLG, which reveals two sets of kagome bands, namely KB1/DP1 (in green) and KB2/DP2 (in brown). These six bands represent characteristic bands for a distorted diatomic kagome lattice with opposite signs of hopping parameters. In the absence of spin-orbit coupling (SOC), the states at the Dirac point of KB1 and at the Γ-point of KB2 and DP2 exhibit small gaps due to inversion symmetry breaking. Additionally, orbital projections further reveal that the kagome bands primarily originate from C $p_z$ orbitals, with minor contributions from in-plane $p$ orbitals, as detailed in Fig. S5.

The real-space wavefunction norm squares $|\varphi|^2$ of these kagome bands further illustrate their intrinsic connection to the kagome lattices. To distinguish the wavefunction distribution properties more clearly, Figs. 3(d-g) present layer resolved $|\varphi|^2$ of the kagome bands in SN-gs-BLG at the K point of the first Brillouin zone (highlighted with solid dots in Fig. 3a). For FB1, the $|\varphi|^2$ in the top-layer (Fig. 3d) is predominantly localized in the AB/BA stacking domains, exhibiting hexagonal lattice symmetry. In contrast, the $|\varphi|^2$ in the bottom-layer (Fig. 3e) is concentrated in the SP stacking domains (Fig. 3e), forming an electronic kagome lattice, thereby confirming

the presence of the kagome flat band in SN-gs-BLG. In contrast, the layer-resolved $|\varphi|^2$ distribution of DP1 (Figs. 3f and 3g) reveals that the state in the top-layer (Fig. 3f) is concentrated within the AA domains, while their interstitial region forms a hexagonal pattern. The state in the bottom-layer (Fig. 3g) exhibits a qualitatively complementary pattern, in which the regions with higher charge densities (shown in darker red) form an obscured electronic kagome lattice. This layer-resolved wavefunction behavior emphasizes the critical interplay between stacking domains and the resulting electronic structure of these kagome bands.

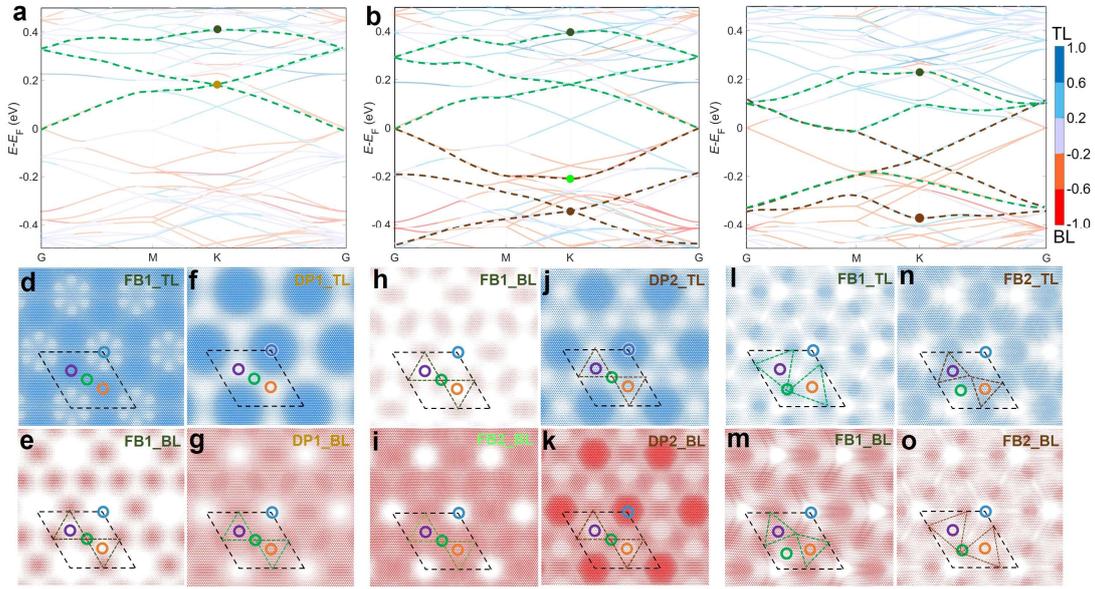

Figure 3 Electronic properties of gs-BLGs. (a-c) Layer atomic projection energy band diagrams for SN-gs-BLG, SC-gs-BLG, and NC-gs-BLG. Gray-blue colored lines represent bands combinedly contributed from both the top-layer and bottom-layer, while the light blue and orange lines indicate dominance by the top-layer and bottom-layer. Darker blue (red) lines, signify nearly exclusive contribution from the top- (bottom) layer. (d, e) The $|\varphi|^2$ of the top-layer (TL) and bottom-layer (BL) for the flat band (dark green dot) in (a), and (f, g) the $|\varphi|^2$ of the TL and BL for the Dirac point (dark gold dot). (h) The $|\varphi|^2$ of the TL for FB1 (dark green) in (b), (i) the $|\varphi|^2$ of the BL for FB2 (bright green), and (j, k) the $|\varphi|^2$ of the TL and BL for DP2 (brown dot). (l, m) the $|\varphi|^2$ of the TL and BL for FB1 (dark green dot) in (c), and (n, o) the $|\varphi|^2$ of the TL and BL for FB2 (brown dot). The isosurface value of the wavefunctions are $1\times10^{-5}$ e Å$^{-3}$. Black dashed diamonds and differently colored circles represent the supercell and the four stacking orders in the same scheme used in Fig. 1a

For SC-gs-BLG (Figs. 3h-3k), the $|\varphi|^2$ in the bottom-layer of FB1 (Fig. 3h) exhibits a similar pattern to that of SN-gs-BLG (Fig. 3e), while the distribution of DP1 is analogous to that in SN-gs-BLG and is therefore not shown. For FB2, the $|\varphi|^2$ in the bottom-layer (Fig. 3i) shows a delocalized feature mapped with an obscured kagome pattern, which is consistent with its larger dispersion compared to that of FB1. The DP2 states projected on the top (Fig. 3j) and bottom (Fig. 3k) layers exhibit similar features in that the wavefunctions are localized around the AA regions. However, the wavefunction observed at the SP regions is complementary in the two layers.

The wavefunction distributions of FB1 and FB2 in NC-gs-BLG (Figs. 3l-3o) are particularly interesting because the distribution of bands on layers is locked. In particular, the projections of FB1 on the top-layer (Fig. 3l) and FB2 on the bottom layer (Fig. 3o) share the same distorted kagome lattice feature. In comparison, FB1 on the bottom-layer (Fig. 3m) and FB2 on the top-layer (Fig. 3n) exhibit another distorted kagome lattice feature. The rotation operations are opposite in direction in these two lattices. Although not strictly identical, these distorted kagome lattices are similar to the geometry and the interlayer-spacing map. These facts indicate that the emergent kagome bands originate from particular distributions of electronic states, driven by structural relaxation-induced local distortions.

Interlayer charge redistributions are also related to the AB and BA stacking domains. Figure S6 displays the differential charge density (DCD) of these three gs-BLGs, illustrating that interlayer charge redistribution, primarily occurring within AB and BA domains, grows progressively stronger under compressive strain. Furthermore, we plotted the electronic structures of SS-gs-BLG and CC-gs-BLG in Fig. S7, where robust Dirac and flat bands are also observed in their band structures. Given these two additional cases, we reveal a trend for bandwidth variation of kagome bands under gradient strains. Figure 4a shows monotonic variations in the bandwidths of the Dirac and flat bands as the lattice shrinks. Specifically, we observe that a pure stretched strain favors more linear Dirac and flatter flat bands. As the degree of compressive strain freedom increases, the energy dispersion of Dirac and flat bands converges. This behavior suggests a potential analogy to the angle dependence observed in twisted bilayer graphene[47]. By applying appropriate gradient strain over larger-sized samples, the Dirac bands near the Fermi level may potentially be fully flattened, paving the way for exotic correlated electronic states.

**Correlation of Electronic States and Out-of-plane Distortions**

Next, we establish the correlation between the kagome electronic states and the out-of-plane structural distortion. Here, the out-of-plane distortions are quantified by projected lengths along the z-axis of C-C bonds (see Fig. S4 for details). Positive and negative projection values indicate that that the distortion-induced atomic displacements are aligned and the z-axis is anti-aligned. Although the out-of-plane distortion is negligible (<0.01 Å) in SN-gs-BLG, the zero out-of-plane displacement regions (white in Fig. 4b) already form a kagome lattice, as highlighted with red ellipses. Each zero-displacement region spans the two smallest interlayer-spacing regions marked in Fig. 2d, exhibiting its spatial anisotropy in the real-space. Moreover, these regions are also spatially corresponding to the observed charge density of kagome electronic states shown in Figs. 3e and 3g.

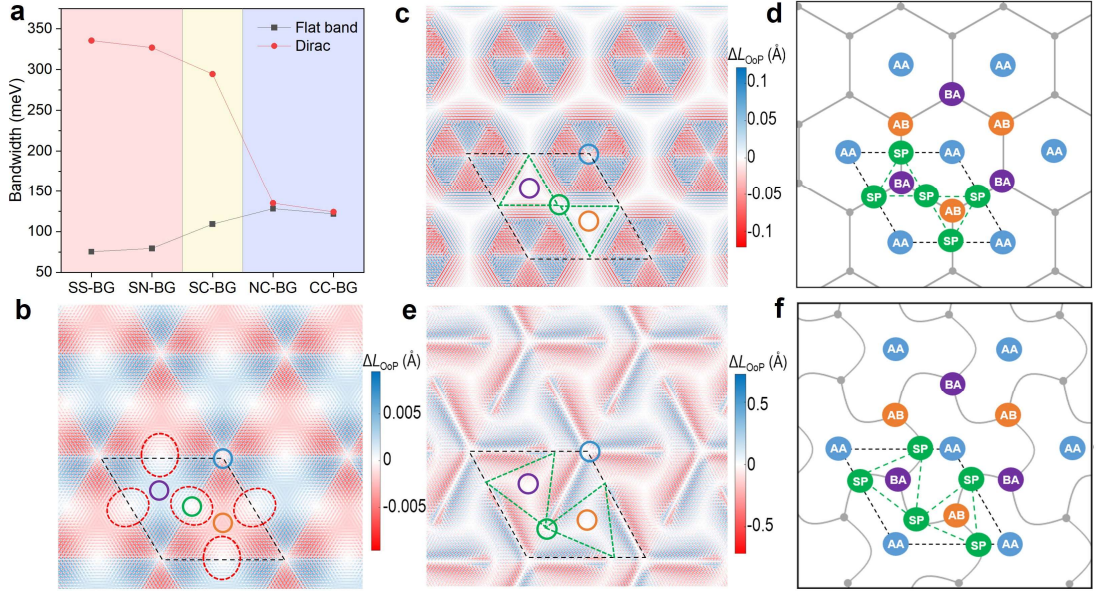

Figure 4 (a) Energy dispersion of kagome bands in five gs-BLGs. (b) Out-of-plane distortion for top-layer of SN-gs-BLG. (c) Out-of-plane distortion for the bottom-layer of SC-gs-BLG and (d) the illustration of kagome line graph framework. (e) Out-of-plane distortion for the bottom-layer of NC-gs-BLG and (f) the illustration of distorted kagome lattice line graph framework, where the cores (grey) are located on the honeycomb sublattice and the linkers (green) on the distorted kagome sublattice. Black dashed diamonds and differently colored circles represent the supercell and the four stacking orders in the same scheme used in Fig. 1a

The introduction of compression in SC-gs-BLG leads to larger out-of-plane displacements (>0.10 Å), as shown in Fig. 4c. Unlike the SN-gs-BLG case, the zero-displacement regions are connected into a continuous hexagonal pattern, which covers all smallest interlayer-spacing regions spatially. This hexagonal pattern is an analogue to the hexagon lattice shown by gray lines in Fig. 4d. The wavefunctions are visualized in Figs. 3h and 3i indicate the kagome electronic states are distributed around the middle points of the hexagonal edges (the SP regions). These SP regions represent the nodes of the line graph of the original hexagonal lattice, in other words, a kagome lattice.[48]

The out-of-plane displacements further enlarge to over 0.5 Å in NC-gs-BLG and form a chiral pattern of the out-of-plane zero-displacement regions (white in Fig. 4e), sufficiently relieving in-plane compressive strains. This chiral zero-displacement pattern covers most of the rotated AB and BA domains, consistent with the distorted diatomic kagome lattice observed in the strongly varied interlayer-spacing distribution map. It is also in analogue to anti-trichiral structures already used in mechanistic metamaterials[49-52], as reproduced by gray curved lines shown in Fig. 4f. Like those in SC-gs-BLG, the kagome electronic states are also distributed around the SP regions. Nevertheless, these SP regions converge to the AA regions, not at the middle points of the AB and BA regions, unlike those in SC-gs-BLG. The line graph (Fig. 4f) of the chiral zero-displacement pattern outlines a distorted kagome lattice where the SP regions (green in Fig. 4f) are located at the nodes of the lattice. Moreover, Fig. S8 depicts the maps of local strains for SS- and CC-gs-BLGs.

## Discussion

These findings indicate that the interlayer coupling drives out-of-plane structural distortions under in-plane gradient strain, giving rise to three different diatomic kagome patterns. The SN-gs-BLG and SC-gs-BLG primarily release in-plane strain differences through in-plane distortions, sufficient to accommodate the imposed strains. As a result, out-of-plane distortions remain minimal, favoring the interlayer interactions while maintaining the same symmetry as the in-plane distortions, ultimately forming undistorted diatomic kagome lattices. In NC-gs-BLG, however, the in-plane compressive strain is significantly large, making it unlikely to be fully released through in-plane distortions alone. Instead, the out-of-plane distortions dominate, nearly two orders of magnitude larger than the in-plane ones, which compete with interlayer stacking interactions, forming of a chiral distorted kagome lattice. The emergence of these strain-induced distortions provides strong evidence for the realization of kagome lattice structures in fully relaxed gs-BLGs.

In summary, we have demonstrated that untwisted bilayer graphene under gradient strains (gs-BLG) can form moiré superlattices that exhibit kagome electronic states. Through detailed force-field calculations for structural relaxations and first-principles calculations for electronic properties, we established that the competition between interlayer coupling and in-plane strain relaxation governs the structural deformations, ultimately leading to the formation of three diatomic kagome lattices under different gradient-strain conditions. The electronic structure of these strained bilayers reveals kagome bands featuring Dirac and flat bands, with their bandwidths tunable via strain modulation. Notably, we observed that compressive strain enhances out-of-plane distortions, which not only significantly influence the electronic states but also induce structural chirality, potentially enabling further tailoring of kagome-related electronic properties. As a feasible tool for engineering electronic structures in 2D materials, the introduction of gradient strain provides a robust and controllable approach to constructing moiré superlattices without the limitations imposed by twist-induced moiré disorders. Our results suggest a promising avenue for the experimental exploration of strain-engineered correlated electronic states in untwisted graphene bilayers and beyond, with implications for novel quantum states.

**Methods**

All structural relaxations of the homogenously strain graphene bilayers were performed using the implementation of a force field, facilitated by the Large Atomic/Molecular Massively Parallel Simulator (LAMMPS)[53]. The C-C interactions within each graphene layer were described using a many-body Tersoff potential, while the interlayer van der Waals interactions were accounted for via the Kolmogorov-Crespi (KC) potential[54,55]. All atoms were fully relaxed until the residual force per atom was less than $1.0 \times 10^{-4}$ eV/Å. The lattice constant of the graphene primitive cell calculated by LAMMPS is 2.46 Å, in agreement with the experiment. The local in-plane strain is defined as $\epsilon_{IP}(r) = \frac{l(r)\cos[\alpha(r)] - l_{CC}^0}{l_{CC}^0}$, where $r$ represents the position coordinates of a C-C bond in the fully relaxed structure, $l$ is the fully relaxed length of the C-C bond, $l_{CC}^0 = 1.42$ Å is the equilibrium bond length of freestanding graphene, and $\alpha$ denotes the angle between the bond and the x-y plane. For simplify, the in-plane strain is visualized through in-plane geometrical displacement $\Delta L_{IP}(r) = \epsilon_{IP}(r) l_{CC}^0$ in Supplementary Fig. 4. Analogously, the local out-of-plane (shear) strain is formulated as $\epsilon_{Oop}(r) = \frac{l(r)\sin[\alpha(r)]}{l_{CC}^0}$. Its corresponding out-of-plane displacement is defined as $\Delta L_{OoP}(r) = \epsilon_{Oop}(r) l_{CC}^0$, providing a direct metric of out-of-plane distortion.

Density functional theory calculations were performed using the generalized gradient approximation (GGA) with the Perdew-Burke-Ernzerhof (PBE) for the exchange-correlation potential, linear combined atomic orbitals (LCAO) method, and a single-zeta polarized (SZP) atomic orbital basis set as implemented in the RESCU package[56]. The real space mesh resolution was set to 0.35 Bohr, and the convergence criteria for electronic energy and charge density are both set to $10^{-5}$ to ensure good convergence. Two *k*-meshes of $3 \times 3 \times 1$ and $1 \times 1 \times 1$ were adopted for calculations on the $11 \times 11 + 12 \times 12$ and $35 \times 35 + 36 \times 36$ supercells, respectively. The models have a vacuum thickness up to about 36 Å in the z-direction to avoid interactions due to periodicity at the surface. For the layer-resolved wavefunction, we obtain the bilayer partial charge density at the K point of the first Brillouin zone, subtract the total charge density of either the top or bottom layer, and visualize the charge accumulation region to represent the layered wavefunction.

**Acknowledges:** We thank Drs. Kui Gong, Yibin Hu, and Yin Wang (all from HZWTECH) for helpful discussions. We gratefully acknowledge the financial support from the National Natural Science Foundation of China (Grants No. 52461160327, No. 92477205, No. 12474173 and No. 12104313), the National Key R&D Program of China (Grant No. 2023YFA1406500), the Research Grants Council of Hong Kong (Grant No. CRS_PolyU501/24), the Department of Science and Technology of Guangdong Province (No. 2021QN02L820) and Shenzhen Science and Technology Program (Grant No. RCYX20231211090126026, the Stable Support Plan Program 20220810161616001), the Fundamental Research Funds for the Central Universities, and the Research Funds of Renmin University of China (Grants No. 22XNKJ30). Calculations were performed at the Physics Lab of High-Performance Computing of Renmin University of China, PCC@RUC.